# GAUSSIAN SPEAKER EMBEDDING LEARNING FOR TEXT-INDEPENDENT SPEAKER VERIFICATION

*Bin Gu, Wu Guo*

National Engineering Laboratory for Speech and Language Information Processing,
University of Science and Technology of China, Hefei, China

## ABSTRACT

The x-vector maps segments of arbitrary duration to vectors of fixed dimension using deep neural network. Combined with the probabilistic linear discriminant analysis (PLDA) backend, the x-vector/PLDA has become the dominant framework in text-independent speaker verification. Nevertheless, how to extract the x-vector appropriate for the PLDA backend is a key problem. In this paper, we propose a Gaussian noise constrained network (GNCN) to extract x-vector, which adopts a multi-task learning strategy with the primary task classifying the speakers and the auxiliary task just fitting the Gaussian noises. Experiments are carried out using the SITW database. The results demonstrate the effectiveness of our proposed method.

## 1. INTRODUCTION

Speaker verification (SV) is the task of verifying a person's claimed identity from some speech signal. SV systems typically consists of two main stages: (1) a frontend that converts a variable-length utterance to a low- and fixed-dimensional vector, and (2) a backend for calculating the similarity between speaker representations. For the past decade, the combination of i-vector [1] and probabilistic linear discriminant analysis (PLDA) [2] has become the state-of-the-art approach in the SV field.

Since the great success of deep learning over a wide range of machine learning tasks, more attention has been drawn to the use of deep neural network (DNN) to generate speaker vectors having more discriminative power. In most deep speaker embedding systems, a frame-level feature extractor is designed firstly, which can be modeled by convolution neural network (CNN) [3, 4], time-delay neural network (TDNN) [5, 6], recurrent neural network (RNN) [7] or their variants [8, 9, 10]. Next, a pooling layer is exploited to reduce the temporal dimension of frames-level features to get a fixed-dimensional vector and the speaker representation is generated from the following stacked fully connected layers. Typically, statistics pooling [5] combined with different attention mechanisms [11, 12, 13] is used to replace the average pooling for capturing long-term speaker characteristic more effectively.

As we all know, a speech contains a lot of information (such as phoneme, emotion or noise) and the speaker identity is weak information. How to extract more robust and discriminative speaker embedding is always a research focus. Recently, many deep speaker embedding systems, which have a primary task of classifying the target speakers and an auxiliary task, have been proposed. Some researchers find the high-order statistics [14] and phonetic labels [15, 16] of the input acoustic features are helpful for training the model. Furthermore, SNR values, the labels of environment types [17], channel types [18] or language types [19, 20] of utterances are also utilized in some systems to minimize the domain mismatch between the training data and test data. In a word, all these methods are implemented using multi-task learning frameworks.

However, a potential problem of most deep learning methods is that there is a mismatch of training loss and LDA/PLDA training objectives, as noticed in [21]. In order to make the embedding output suitable for the backend, a Gaussian-constrained training method is proposed [22]. The strategy is to minimize intra-class variations, so the performance is highly dependent on the center vectors of different speakers. In addition, the additive regularization is not added on all the speaker embedding layers, which weakens the ability of the final embeddings to fit the Gaussian distribution.

In this work, we propose a Gaussian noise constrained network (GNCN) which adopts multi-task learning framework to extract speaker representations. As the auxiliary task, normal distributed noise vectors will be fitted in the embedding layers. In this way, the secondary objective can make the distribution of all the x-vectors to fit Gaussian distribution. To the best of our knowledge, no study has yet been done on deep speaker embedding from such a perspective. We evaluated our experiments on the SITW evaluation dataset. The experimental results show the proposed methods can improve the performance of the DNN embedding system.

The remainder of this paper is organized as follows. Section 2 describes the related works, including the x-vector baseline system and Gaussian-constrained training algorithm. Section 3 introduces the proposed method. The experimental set up, results and analysis are presented in Section 4. Finally, the conclusion is given in Section 5.

## 2. RELATED WORKS

### 2.1. X-vector baseline system

The network architecture of our x-vector baseline system is the same as that described in [6]. Five TDNN (or 1-dimensional dilated CNN) layers $l_1$ to $l_5$ are stacked for extracting the frame-level features. More specifically, the second and third layers with dilated filters are exploited to efficiently enlarge the receptive-field size with low computation complexity, while the others retain the dilation rate of 1. The kernel sizes of the five layers are 5, 3, 3, 1 and 1, respectively.

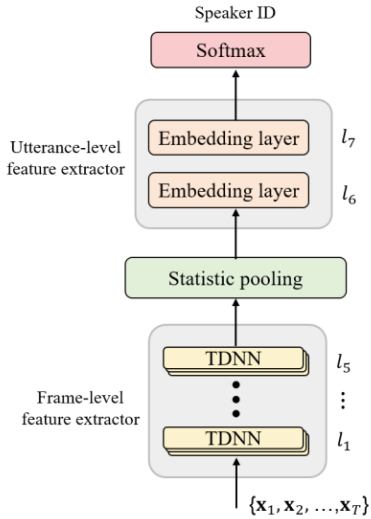

**Figure 1:** *Network architecture of the x-vector baseline.*

The final frame-level output vectors of the whole variable length utterance are aggregated into a fixed segment-level vector through the statistics pooling layer. The mean and standard deviation are both calculated and then concatenated together as the output of the statistics pooling layer. Two additional fully connected layers $l_6$ and $l_7$ are added to obtain a low-dimensional utterance-level representation that is finally passed into a softmax output layer. Each of its output nodes corresponds to one speaker ID and cross entropy (CE) loss function is used.

Once the DNN is trained, we remove the softmax layer and the last fully connected layer $l_7$, and the output of the first linear affine layer directly on top of the statistics pooling is extracted as the speaker embedding.

### 2.2. Gaussian-constrained training

Gaussian-constrained training algorithm [22] urges the model producing Gaussian distributed speaker vectors. More specifically, a regularization term, which constrains the output distribution of the network, is added to the training objective.

Suppose $f(\mathbf{x})$ is the x-vector of utterance $\mathbf{x}$. A little different from the softmax classifier, each speech utterance $\mathbf{x}$ is classified as follows:

$$p(s \mid f(\mathbf{x})) = \frac{e^{f(\mathbf{x}) \cdot \mathbf{\theta}_s}}{\sum_{s'} e^{f(\mathbf{x}) \cdot \mathbf{\theta}_{s'}}} \quad (1)$$

where $\mathbf{\theta}_s$ represents the parameters in the classifier that are associated with the output node corresponding to speaker $s$. Then the CE loss can be calculated. In addition, a regularization term $\mathcal{R}$ is designed to control the distribution of $f(\mathbf{x})$:

$$\mathcal{R} = \sum_s \sum_{x \in \varepsilon(s)} \| f(\mathbf{x}) - \mathbf{\theta}_s \|_2 \quad (2)$$

where $\varepsilon(s)$ is the set of utterances belonging to speaker $s$. Finally, the training objective, which is composed of the CE loss and the regularization $\mathcal{R}$, will be used for training the model. In order to control the strength of $\mathcal{R}$, it is multiplied by a weight $\alpha$.

From Eq.2, we can see that the regularization term just encourages all the utterance-level x-vectors belonging to speaker $s$ to converge to a center vector $\theta_s$, but the prior of the whole x-vectors is not constrained.

## 3. PROPOSED METHOD

The architecture of the proposed GNCN is depicted in Figure.2. The primary task is the same as the x-vector baseline, whereas an auxiliary task is to fit Gaussian distributed noise vectors. From another perspective, the primary task is to train a discriminative model (yellow block) and auxiliary task is to obtain a generative model (blue block). Our goal is to minimize the distance between the distributions of these two models, so we can make both models sharing most parameters.

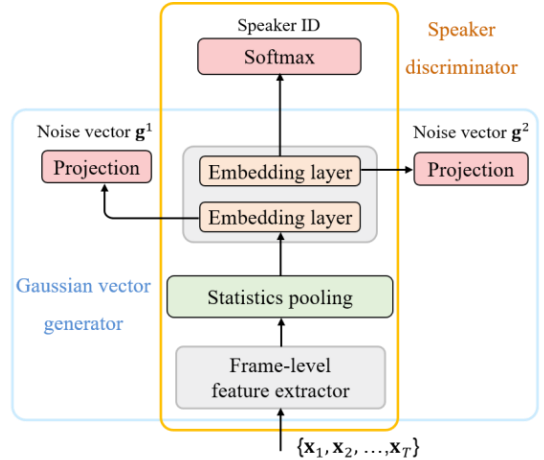

**Figure 2:** *The architecture of the proposed GNCN*

We train the classification model using a standard CE loss. Meanwhile, we aim to minimize the mean square error (MSE) loss in the generation model between its output and noise vectors. Training with the augmented objective will encourage the model to produce speaker representations which are more suitable for the backend classifier.

Suppose there are $N$ samples in each batch and $L$ utterance-level embedding layers. We denote $n^{th}$ x-vector extracted from the $l^{th}$ embedding layer by $\mathbf{v}_n^l$. The linear transformed speaker vector through a projection will be used as the output of the generation model, so the MSE loss can be formulated as follow:

$$\tilde{\mathbf{v}}_n^l = \mathbf{W}^l \mathbf{v}_n^l + \mathbf{b}^l \quad (3)$$

$$\mathcal{L}_{MSE} = \frac{1}{N} \sum_{l=1}^{L} \sum_{n=1}^{N} \left\| \tilde{\mathbf{v}}_n^l - \mathbf{g}_n^l \right\|_2 \quad (4)$$

where $\mathbf{W}^l$, $\mathbf{b}^l$ and $\mathbf{g}_n^l$ are the weight, bias parameters and the Gaussian noises respectively. Combined with the original CE loss, the final loss function can be written as follow:

$$\mathcal{L} = \mathcal{L}_{CE} + \lambda \mathcal{L}_{MSE} \quad (5)$$

where the $\lambda$ is the task weight. Note, $\mathcal{L}_{MSE}$ will converge to a certain value quickly, and therefore $\lambda$ should be decayed based on the validation set to ensure the auxiliary task working but not playing a dominant role.

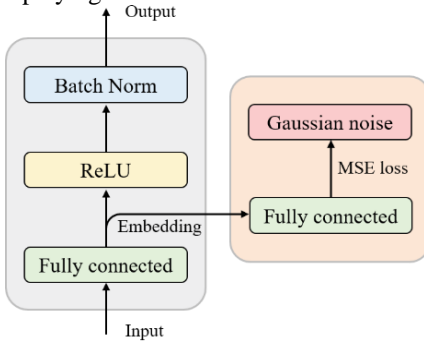

**Figure 3:** *Structure details of an embedding layer and the two layers in our systems have the same structure.*

We also try other GNCN structure. As shown in Figure.3, a linear affine layer is appended to the output of the fully connected layer before calculating the MSE loss. The noise vectors are low-dimensional, therefore adding multiple linear layers only slightly increases the number of parameters. Moreover, the extra overhead can be neglected, since the top layers are removed when extracting speaker embeddings. Fig.3 just depicts the flowchart of one layer, and both the utterance level layers adopted the proposed method. Through introducing the regularization term into higher layer, the auxiliary task can also constrain the prior of the x-vectors in non-linear space.

When the model converges, speaker representations are extracted from the first embedding layer. They can be viewed as the Gaussian vector obtained from the generation model with the discriminative speaker information captured from the classification model.

## 4. EXPERIMENTS
### 4.1. Data set and evaluation metric

All the experiments are conducted on the SITW database [23]. There are two standard datasets for testing: Dev. Core and Eval. Core. We use both these two sets to conduct the experiments. The VoxCeleb database [24], including the VoxCelebb1 and VoxCeleb2, is used for training. Since a few speakers are included in both SITW and Voxceleb, these speakers are removed from the training dataset. Due to the background noise such as laughter and music sampled from the real world, data augmentation techniques described in [6] including adding additive noise and reverberation data are applied to improve the robustness of the system.

The results are reported in terms of three metrics: the equal error rate (EER), and the minimum of the normalized detection cost function (minDCF) with two settings: one with the prior target probability $P_{tar}$ set to 0.01 (DCF($10^{-2}$)), and the other with $P_{tar}$ set to 0.001 (DCF($10^{-3}$)).

### 4.2. Features

30-dimensional MFCC features extracted from the speech of a 25ms window with 10ms frame shift are used. They are mean-normalized over a 3 second sliding window, and energy based VAD is employed to filter out non-speech frames. The acoustic features are randomly cropped to lengths of 2-4s, and 128 utterances with the same duration are grouped into a mini-batch. The data processing is implemented with the Kaldi toolkit [26].

### 4.3. Model configuration

The deep embedding network is implemented using the Tensorflow toolkit [27]. We use the Adam optimizer with an initial learning rate of 0.001 and reduce it to 0.0001 gradually. The same type of batch normalization and L2 weight decay as described in [25] are used to prevent overfitting. Except where specifically noted, all the setups are the same with the baseline system. We mainly compare five systems, and several variants of the proposed system are also explored.

**x-vector**: This is the baseline system, the configuration is the same as that introduced in Section 2.1. Each of the first four frame-level hidden layers has 512 nodes, while there are 1536 hidden nodes in the fifth layer. Both of the two fully connected layer after statistics pooling layer have 512 nodes. The nonlinear activation function of each hidden layer is ReLU.

**GTM**: The Gaussian-constrained training method described in Section 2.2 is applied in this system. The task weight is set to 0.05 which keeps the same with that in [22].

**GNCN-F0/F1-FC**: These are the proposed systems. In the systems with the prefix "GNCN-F0", the raw embeddings are directly used for calculating the MSE loss with noise vectors, while the others use the linear transformed speaker vectors through one affine layer with 100 hidden nodes. The hyper-parameter $\lambda$ is empirically set to 0.1. Similar to the structure depicted in Figure.3, the auxiliary task can be added to other positions of the main network. The suffixes "IN", "FC", "AF" and "BN" denote that the branches of the auxiliary task are directly added to the input, the hidden vectors after fully connected layer, activation function and batch norm, respectively.

Table 1 *Results of different systems on SITW*

| system | Dev | | | Eval | | |
|---|---|---|---|---|---|---|
| | EER(%) | DCF($10^{-2}$) | DCF($10^{-3}$) | EER(%) | DCF($10^{-2}$) | DCF($10^{-3}$) |
| x-vector | 2.757 | 0.2918 | 0.4834 | 3.226 | 0.3304 | 0.5486 |
| GTM | 2.580 | 0.2730 | 0.4537 | 3.198 | 0.3150 | 0.4968 |
| GNCN-F0-FC | 2.595 | 0.2714 | **0.4196** | 2.962 | 0.2984 | **0.4628** |
| GNCN-F1-FC | 2.441 | 0.2798 | 0.4447 | 2.798 | 0.3037 | 0.4871 |
| GNCN-Fusion | **2.349** | **0.2653** | 0.4242 | **2.661** | **0.2948** | 0.4810 |

Table 2 *Comparison results of applying the proposed method in different positions of embedding layer*

| system | Dev | | | Eval | | |
|---|---|---|---|---|---|---|
| | EER(%) | DCF($10^{-2}$) | DCF($10^{-3}$) | EER(%) | DCF($10^{-2}$) | DCF($10^{-3}$) |
| GNCN-F1-IN | 2.580 | 0.2896 | 0.4725 | 3.198 | 0.3159 | 0.4910 |
| GNCN-F1-FC | **2.441** | **0.2798** | **0.4447** | **2.798** | **0.3037** | **0.4871** |
| GNCN-F1-AF | 2.888 | 0.3090 | 0.4854 | 3.253 | 0.3397 | 0.5471 |
| GNCN-F1-BN | 2.707 | 0.2961 | 0.4902 | 3.144 | 0.3295 | 0.5589 |

**GNCN-Fusion**: The complementarity between the GNCN-F0-FC and GNCN-F1-FC is also investigated here. We only report the results using the score fusion of the GNCN-F0 and GNCN-F1 with equal weights.

### 4.4. Backend classifier

After extracting x-vectors, the evaluation set are centered using the training set. The dimensions of the vectors are reduced to 100 through LDA algorithm. Length normalization is adopted before PLDA. After these pre-processing steps, the PLDA model is trained and used as backend classifier for speaker verification.

### 4.5. Results and analysis

Table 1 presents the results of different systems on SITW. It can be observed that all the systems with constrained learning methods outperform the x-vector baseline system. Furthermore, our proposed GNCNs outperform the GTM and x-vector baseline in all evaluation conditions. With regard to the results of GNCNs, there is an interesting finding. GNCN-FC-F0 performs best in terms of DCF and improves the DCF($10^{-3}$) by 16% on the evaluation set compared with the baseline system, while GNCN-FC-F1 achieves best result of single system in terms of EER. In fact, the former system can be viewed as a special case of the latter system. Among the above systems, the fused system achieves 18% and 11% relative improvements compared with the baseline system in terms of the EER and DCF($10^{-2}$) respectively. This result makes clear that the TDNN-based x-vector system is enhanced significantly with our introduced method.

In the systems listed in the table 1, the auxiliary task is added to the hidden nodes after fully connected layer. In the following experiments, we investigate the effect of the position where the auxiliary task is introduced. Since we can obtain similar results for the GNCN-F0 systems, Table 2 only lists the results of the GNCN-F1 systems. It can be easily found that GNCN-F1-FC gets the best performance among the four systems, which demonstrates that directly constraining the speaker representations is more effective. Moreover, GNCN-F1-AF and GNCN-F1-BN just get comparable or even worse results compared with the baseline system in Table 1. Both of them just introduce the regularization term on the non-linear transformed speaker embeddings, and this phenomenon means that only fitting Gaussian noises in non-linear space may not work very well.

## 5. CONCLUTION

In this paper, we proposed a Gaussian noise constrained x-vector extraction scheme that can generate better speaker embeddings for the backend classifier. More specifically, an auxiliary task of fitting noise vectors is adopted in each embedding layer. With the normal distribution constrained, the distribution of the x-vectors can be more like a Gaussian. The experimental results demonstrate significant performance gains over the conventional methods. In the future, we will incorporate the noise vectors and adversarial training algorithms which have achieved great success in domain adaption, and conduct more comprehensive analysis.

## 6. ACKNOWLEDGEMENTS

This work was partially funded by the National Natural Science Foundation of China (Grant No. U1836219) and the National Key Research and Development Program of China (Grant No. 2016YFB100 1303).